\documentclass{elsart5p}

\usepackage{graphicx}
\begin{document}

\begin{frontmatter}
\title{Diatomic molecule as a quantum entanglement switch}
\author{Adam Rycerz\corauthref{Rycerz}}
\ead{rycerz@th.if.uj.edu.pl}
\address{
  Marian Smoluchowski Institute of Physics, Jagiellonian University, 
  Reymonta 4, 30--059 Krak\'{o}w, Poland
}
\corauth[Rycerz]{
  Corresponding author. Tel: (+48 12) 663--55--68  Fax: (+48 12) 633--40--79
}

\begin{abstract}
We investigate a pair entanglement of electrons in diatomic molecule, modeled 
as a correlated double quantum dot attached to the leads. The low-temperature 
properties are derived from the ground state obtained by utilizing the 
Rejec-Ram\v{s}ak variational technique within the framework of EDABI method, 
which combines exact diagonalization with \emph{ab initio} calculations. The 
results show, that single-particle basis renormalization modifies the 
entanglement-switch effectiveness significantly. We also found the entanglement
signature of a competition between an extended Kondo and singlet phases.
\end{abstract}


\end{frontmatter}

Quantum entanglement, as one of the most intriguing features of quantum 
mechanics, have spurred a great deal of scientific activity during the last 
decade, mainly because it is regarded as a valuable resource in quantum 
communication and information processing \cite{qinfo}. The question on 
entanglement between microscopic degrees of freedom in a condensed phase have 
been raised recently \cite{spinc}, in hope to shed new lights on the physics 
of quantum phase transitions and quantum coherence \cite{webza}. In the field 
of quantum electronics, a pair entanglement appeared to be a convenient tool 
to characterize the nature of transport through quantum dot, since its vanish 
when the system is in a Kondo regime \cite{rycent}. The analogical behavior was
observed for two qubits in double quantum dot, for either \emph{serial} and 
\emph{parallel} configuration \cite{ramra}. The latter case is intriguing, 
since the concurrence \cite{woot} at $T=0$ changes abruptly from $\mathcal{C}
\approx 1$ to $\mathcal{C}=0$ when varying the interdot coupling, so a finite 
Anderson system shows a true quantum phase transition.

\begin{figure}[!t]
\begin{center}
\includegraphics[width=\columnwidth]{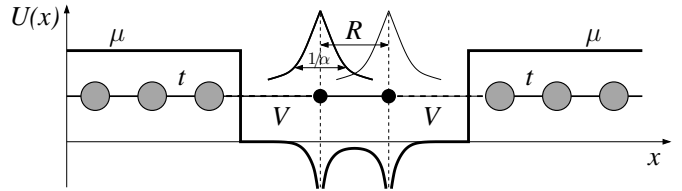}
\end{center}
\caption{
 Diatomic molecule modeled as a double quantum dot attached serially to the 
leads. A cross-section of the single-particle potential along the main system 
axis is shown schematically.} \label{qd2fig}
\end{figure}

Here we consider a nanoscale version of such an \emph{entanglement switch}, 
inspired by conductance measurements for a single hydrogen molecule 
\cite{smit}. A special attention is payed to electron-correlation effects, in 
particular the wave-function renormalization \cite{spanew}. Recent experiment 
\cite{djuk} shows the current through a molecule is carried by a single 
conductance channel, so serial configuration shown in Fig.\ \ref{qd2fig} seems 
to be the realistic one. The Hamiltonian of the system is
\begin{equation}
\label{ham5}
  H = H_L+V_L+H_C+V_R+H_R,
\end{equation}
where $H_C$ models the central region, $H_{L(R)}$ describes the left (right) 
lead, and $V_{L(R)}$ is the coupling between the lead and the central region. 
Both $H_{L(R)}$ and $V_{L(R)}$ terms have a tight--binding form, with the 
chemical potential in leads $\mu$, the hopping $t$, and the tunneling amplitude
$V$, as depicted schematically in~Fig.\ \ref{qd2fig}. The central-region 
Hamiltonian 
\begin{equation}
\label{hamc}
  H_C = \sum_{ij\sigma}t_{ij}c_{i\sigma}^{\dagger}c_{j\sigma}+\sum_{i\sigma %
  \neq j\sigma'}U_{ij}n_{i\sigma}n_{i\sigma'} + (Ze)^2/R
\end{equation}
(with $i,j=1,2$ and $\sigma=\uparrow,\downarrow$) describes a double quantum 
dot with electron-electron interaction. $t_{ij}$ and $U_{ij}$ are 
single-particle and interaction elements, the last term describes the Coulomb 
repulsion of the two ions at the distance $R$. Here we put $Z=1$ and calculate 
all the parameters $t_{ij}$, $U_{ij}$ as the Slater integrals \cite{slat} for 
$1s$-like hydrogenic orbitals $\Psi_{1s}(\mathbf{r})=\sqrt{\alpha^3/\pi}
\exp(-\alpha|\mathbf{r}|)$, where $\alpha^{-1}$ is the orbital size 
(\emph{cf.}\ Fig.\ \ref{qd2fig}). The parameter $\alpha$ is optimized to get a 
minimal ground-state energy for whole the system described by the Hamiltonian 
(\ref{ham5}). Thus, following the idea of EDABI method \cite{spanew}, we 
reduce the number of physical parameters of the problem to just a three: the 
interatomic distance $R$, the lead-molecule hybridization $\Gamma=V^2/t$, and 
the chemical potential $\mu$ (we put the lead hopping $t=1\ \mathrm{Ry} = 
13.6\ \mathrm{eV}$ to work in the wide--bandwidth limit).

\begin{figure}[!t]
\begin{center}
\includegraphics[width=0.8\columnwidth]{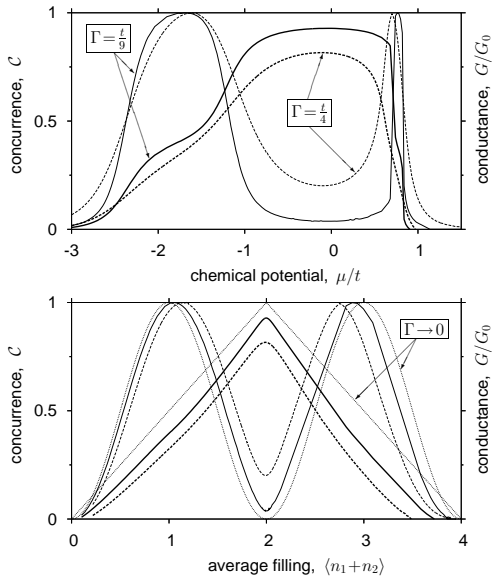}
\end{center}
\caption{
Entanglement and transport through the system in Fig.\ \ref{qd2fig} as a 
function of the chemical potential $\mu$ (\emph{top panel}) and the average 
filling $\langle n_1\!+\!n_2\rangle$ (\emph{bottom panel}). Tick (thin) 
\emph{solid} and \emph{dashed} lines shows the concurrence $\mathcal{C}$ 
(conductance $G$) for $\Gamma=t/9$ and $t/4$, respectively. The limits 
$\Gamma\rightarrow 0$ are depicted with dotted lines in the bottom panel. The 
interatomic distance is $R=1.5a$. } \label{gH2fig}
\end{figure}

\begin{figure}[!t]
\begin{center}
\includegraphics[width=0.8\columnwidth]{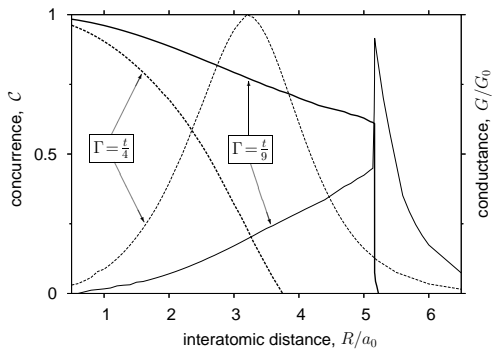}
\end{center}
\caption{
Concurrence (\emph{tick} lines) and conductance (\emph{thin} lines) at the 
half-filled sector $\langle n_1\!+\!n_2\rangle=2$ as a function of the 
interatomic distance R. The remaining parameters are the same as in Fig.\ 
\ref{gH2fig}.} \label{gH2Rfig}
\end{figure}

The entanglement between electrons placed on two atoms can be characterized by 
the charge concurrence \cite{rycent}
\begin{equation}
\label{cconc}
\mathcal{C}=2\max\left\{0,|\langle c_{i\sigma}^{\dagger}c_{j\sigma}\rangle |-
\sqrt{\langle n_{i\sigma}n_{j\sigma}\rangle\langle\bar{n}_{i\sigma}\bar{n}_{ %
j\sigma}\rangle }\right\}
\end{equation}
where $\bar{n}_{i\sigma}\equiv 1\!-\!n_{i\sigma}$. We also discuss the 
conductivity calculated from the formula $G=G_0\sin^2(E_+-E_-)/4tN$ 
\cite{reram}, where $G_0=2e^2/\hbar$, and $E_\pm$ are the ground-state energies
of the system with periodic and antiperiodic boundary conditions, respectively.
Either the energies $E_\pm$ or correlation functions in Eq.\ (\ref{cconc}) are 
calculated within the Rejec--Ram\v{s}ak variational method \cite{reram}, 
complemented by the orbital size optimization, as mentioned above. We use up 
to $N=10^4$ sites to reach the convergence.

In Fig.\ \ref{gH2fig} we show the concurrence and conductance for $R=1.5a_0$ 
(where $a_0$ is the Bohr radius) and two values of the hybridization 
$\Gamma=t/9$ and $t/4$. The conductance spectrum asymmetry, caused by 
wave-function renormalization \cite{spanew}, is followed by an analogical 
effect on entanglement, which changes significantly faster for the upper 
conduction band, where the average filling is $\langle n_1\!+\!n_2\rangle
\approx 3$ (one \emph{extra electron}). The asymmetry vanish when analyzing 
the system properties as a function of $\langle n_1\!+\!n_2\rangle$, showing 
it originates from varying charge compressibility $\chi_c=\partial\langle n_1
\!+\! n_2\rangle/\partial\mu\approx 2/(U_{11}+U_{12})\sim 1/\alpha$. We also 
note the convergence of discussed quantities with $\Gamma\rightarrow 0$ to 
$\mathcal{C}\approx 1-|\langle n_1\!+\! n_2\rangle-2|/2$ and $G\approx G_0
\sin^2(\pi\langle n_1\!+\!n_2\rangle/2)$.

Entanglement evolution with $R$ is illustrated in Fig.\ \ref{gH2Rfig}, where we
focus on the charge neutral section $\langle n_1\!+\!n_2\rangle=2$. The abrupt 
entanglement drop follows the sharp conductance peak for $\Gamma=t/9$, which is
associated with the competition between double Kondo and spin/charge singlet 
phases \cite{mrav}. For $\Gamma=t/4$ both $\mathcal{C}$ and $G$ dependence on 
$R$ become smooth, but the switching behavior is still present. Earlier, we 
have shown that $\Gamma=t/4$ is large enough to cause molecule instability and 
therefore may allow the individual atom manipulation \cite{spanew}. 

In conclusion, we analyzed a pair entanglement of electrons in diatomic 
molecule attached serially to the leads. Entanglement evolution with the 
chemical potential speeds up remarkably for the negatively charged system, due 
to electron correlation effects. The switching behavior was also observed when 
changing the interatomic distance.

The work was supported by Polish Science Foundation (FNP), and Ministry of 
Science Grant No.\ 1 P03B 001 29.



\vspace{-2em}
\begin{thebibliography}{99}

\bibitem{qinfo}
See review by C.H.\ Bennet and D.P.\ Divincenzo, 
\textit{Nature} \textbf{404}, 247 (2000);
M.A.\ Nielsen and I.~L.~Chuang, 
\textit{Quantum Computation and Quantum Information}
(Cambridge, 2000).

\bibitem{spinc}
A.\ Osterloh \emph{et al.}, 
\textit{Nature} \textbf{416}, 608 (2002);
T.\ J.\ Osborne and M.\ A.\ Nielsen, 
\textit{Phys.\ Rev.\ A} \textbf{66}, 032110 (2002);
S.--J.\ Gu \emph{et al.}, 
\textit{Phys.\ Rev.\ Lett.} \textbf{93}, 086402 (2004).

\bibitem{webza}
J.\ van Wezel, J.\ van den Brink, J.\ Zaanen, 
\textit{Phys.\ Rev.\ Lett.} \textbf{94}, 230401 (2005);
\texttt{cond-mat/0606140}.

\bibitem{rycent}
A.\ Rycerz, \textit{Eur.\ Phys.\ J.\ B} \textbf{52}, 291 (2006);
S.\ Oh, J.\ Kim, \textit{Phys.\ Rev.\ B} \textbf{73}, 052407 (2007).

\bibitem{ramra}
A.\ Ram\v{s}ak, J.\ Mravlje, R.\ \v{Z}itko, J.\ Bon\v{c}a, 
\textit{Phys.\ Rev.\ B} \textbf{74}, 241305(R) (2006);
R.\ \v{Z}itko, J.\ Bon\v{c}a, \emph{ibid.} \textbf{74}, 045312 (2006).

\bibitem{woot}
W.K.\ Wootters, 
\textit{Phys.\ Rev.\ Lett.} \textbf{80}, 2245 (1998).

\bibitem{smit}
R.H.M.\ Smit \emph{et al.}, \textit{Nature} \textbf{419}, 906 (2002).

\bibitem{spanew}
J.\ Spa{\l}ek \emph{et al.}, \texttt{cond-mat/0610815}.

\bibitem{djuk}
D.\ Djukic, J.M.\ van Ruitenbeek, \textit{Nano.\ Lett.} \textbf{6}, 789 (2006);
M.\ Kiguchi \emph{et al.}, \texttt{cond-mat/0612681}.

\bibitem{slat}
J.\ C.\ Slater, \textit{Quantum Theory of Molecules and Solids}, 
McGraw--Kill (New York, 1963), Vol.\ 1, p.\ 50.

\bibitem{reram}
T.\ Rejec, A.\ Ram\v{s}ak, 
\textit{Phys.\ Rev.\ B} \textbf{68}, 033306 (2003).

\bibitem{mrav}
P.S.\ Cornaglia, D.R.\ Grempel, 
\textit{Phys.\ Rev.\ B} \textbf{71}, 075305 (2005);
J.\ Mravlje, A.\ Ram\v{s}ak, T.\ Rejec, \emph{ibid.} 
\textbf{73}, 241305(R) (2006).

\end{thebibliography}
\end{document}